\documentclass[twocolumn,pra,twocolumn,superscriptaddress]{revtex4}
\usepackage[latin9]{inputenc}
\usepackage{color}
\usepackage{amsmath}
\usepackage{amssymb}
\usepackage{graphicx}
\usepackage{tikz}

\usepackage{mathrsfs} % cal F fidelity
\usepackage{float}
\usepackage{bbm}

\usepackage{epsfig} 
\usepackage{verbatim}
\usepackage{array}
\usepackage{setspace}
\usepackage{soul}
   
\usepackage[unicode=true,
 bookmarks=true,bookmarksnumbered=false,bookmarksopen=false,
 breaklinks=false,pdfborder={0 0 1},backref=false,colorlinks=true]
 {hyperref}
\hypersetup{
 linkcolor=magenta, urlcolor=blue, citecolor=blue, pdfstartview={FitH}, hyperfootnotes=false, unicode=true}

\makeatletter
\@ifundefined{textcolor}{}
{%
 \definecolor{BLACK}{gray}{0}
 \definecolor{WHITE}{gray}{1}
 \definecolor{RED}{rgb}{1,0,0}
 \definecolor{GREEN}{rgb}{0,1,0}
 \definecolor{BLUE}{rgb}{0,0,1}
 \definecolor{CYAN}{cmyk}{1,0,0,0}
 \definecolor{MAGENTA}{cmyk}{0,1,0,0}
 \definecolor{YELLOW}{cmyk}{0,0,1,0}
}

\usepackage{amsfonts}\usepackage{tabularx}\usepackage{dcolumn}\usepackage{bm}\usepackage{graphicx}\usepackage{epstopdf}

\hypersetup{urlcolor=blue}

\makeatother

\newcolumntype{C}[1]{>{\centering\arraybackslash$}p{#1}<{$}}

\begin{document}

\title{Neural-network-designed pulse sequences for robust control of singlet-triplet qubits}

\author{Xu-Chen Yang}
\affiliation{Department of Physics, City University of Hong Kong, Tat Chee Avenue, Kowloon, Hong Kong SAR, China}
\affiliation{City University of Hong Kong Shenzhen Research Institute, Shenzhen, Guangdong 518057, China}
\author{Man-Hong Yung}
%\email{yung@sustc.edu.cn}
\affiliation{Institute for Quantum Science and Engineering and Department of Physics,
South University of Science and Technology of China, Shenzhen 518055, China}
\affiliation{Shenzhen Key Laboratory of Quantum Science and Engineering, Shenzhen 518055, China}
\author{Xin Wang}
\email{x.wang@cityu.edu.hk}
\affiliation{Department of Physics, City University of Hong Kong, Tat Chee Avenue, Kowloon, Hong Kong SAR, China}
\affiliation{City University of Hong Kong Shenzhen Research Institute, Shenzhen, Guangdong 518057, China}
\date{\today}

\begin{abstract}
Composite pulses are essential for universal manipulation of singlet-triplet spin qubits. In the absence of noise, they are required to perform arbitrary single-qubit operations due to the special control constraint of a singlet-triplet qubits; while in a noisy environment, more complicated sequences have been developed to dynamically correct the error. Tailoring these sequences typically requires numerically solving a set of nonlinear equations. Here we demonstrate that these pulse sequences can be generated by a well-trained, double-layer neural network. For sequences designed for the noise-free case, the trained neural network is capable of producing almost exactly the same pulses known in the literature. For more complicated noise-correcting sequences, the neural network produces pulses with slightly different line-shapes, but the robustness against noises remains comparable. These results indicate that the neural network can be a judicious and powerful alternative to existing techniques, in developing pulse sequences for universal fault-tolerant quantum computation.
\end{abstract}

\maketitle

\setstcolor{red}

\section{Introduction}

High-precision manipulation of quantum-mechanical systems is key to the realization of quantum computation, a next-generation technology promised to be much more powerful than present-day computing devices~\cite{NielsenChuang.00}. Worldwide efforts have been devoted to engineering various physical systems as prototype quantum computers in laboratories. Among them, spin qubits in semiconductor quantum dots represent a promising direction, due to the long coherence time and high control fidelities \cite{Petta.05, Bluhm.10b, Barthel.10,Maune.12, Pla.13,Muhonen.14,Kim.14,Kawakami.16}, as well as the scalability \cite{Taylor.05}. While different types of spin qubits have been theoretically proposed \cite{Loss.98, DiVincenzo.00, Levy.02,Shi.12}  and experimentally demonstrated  \cite{Petta.05, Bluhm.10b, Barthel.10,Brunner.11,Maune.12, Pla.13,Muhonen.14,Kim.14,Kawakami.16}, the singlet-triplet qubit is one of the most studied systems, since it is the most accessible spin qubit that can be controlled solely by  electrostatic means \cite{Petta.05, Foletti.09,Bluhm.10c,Maune.12, Shulman.12, Wu.14, Nichol.16}. In particular, arbitrary rotations around the Bloch sphere can be achieved by combinations of $z$-axis rotations, controlled via the Heisenberg exchange interaction \cite{Petta.05}, and $x$-axis rotations, generated from an inhomogeneous Zeeman field \cite{Foletti.09,Bluhm.10c,Brunner.11,Petersen.13,Wu.14}.

Decoherence of a spin qubit occurs through two main channels: the nuclear (or Overhauser) noise \cite{Reilly.08, Cywinski.09a}, and the charge noise \cite{Hu.06, Nguyen.11, Thorgrimsson.16}. The nuclear noise stems from the hyperfine interaction between the qubit and the surrounding nuclear spin bath. The charge noise originates from impurities near the quantum dots, where electrons can hop on and off randomly, shifting the energy levels of the quantum dot system and subsequently causing control inaccuracy.
Some of these errors are being addressed by various techniques including the dynamical Hamiltonian estimation \cite{Shulman.14}, the use of isotope-enriched silicon substrates \cite{Tyryshkin.11,Muhonen.14,Veldhorst.14}, and resonant gating near certain ``sweet spots'' of the energy spectrum \cite{Medford.13,Bertrand.15,Wong.15,Kim.15}. Alternatively, dynamically-corrected gates \cite{Khodjasteh.09,Khodjasteh.10, Khodjasteh.12, Wang.12, Green.12, Kosut.13,Cerfontaine.14} can be employed instead: they are effective for reducing both nuclear and charge noises, and can be extended to a range of platforms. Inspired by the dynamical decoupling technique, which has been  very successful  in NMR quantum control \cite{Uhrig.07}, the key feature of dynamically corrected gates is the self-compensation of noise: control sequences are tailored such that the errors accumulated during different stages of control acquire different signs and eventually get canceled out to the leading order. 

The physical constraints in controlling a singlet-triplet qubit are as follows. One is only allowed to externally control the rotation around the $z$-axis. As the exchange interaction can neither change sign nor become arbitrarily large, the direction and rotating speed are limited. These constriants make traditional dynamically-corrected gates developed for NMR inapplicable and necessitates the development of control protocols specifically for singlet-triplet qubits. In this context, a family of dynamically corrected gates called  \textsc{supcode} has been proposed \cite{Wang.12}.
In a series of papers, we have developed control sequences resilient to both Overhauser and charge noise, for single and two-qubit operations \cite{Kestner.13,Wang.14a,Wang.15}. These pulses represent theoretical control protocols that would allow for universal manipulation of singlet-triplet qubits \cite{HansonBurkard.07} resistant to noises in the leading order. However, constructing the pulse sequences is resource-consuming, which typically requires a multi-dimensional search for real and positive solutions to a set of coupled non-linear equations.

Born from artificial intelligence, machine learning is essentially a set of techniques allowing analyses of enormous amount of data beyond the ability of human being or any enumerative methods previously imagined. It is now among the most active research fields across all sciences \cite{Jordan.15,LeCun.15,Silver.16}. In recent years, machine learning has been vastly successful in solving problems in various aspects of physics. In cosmological and astro-physics, it has been the workhorse to analyze gravitational waves \cite{Biswas.13,LIGO}. In condensed matter, material and quantum physics  \cite{Biamonte.16},  
it has been successfully applied to material design \cite{Kalinin.15},  turbulent dynamics~\cite{Reddy2016a}, Hamiltonian learning~\cite{Wiebe2014c,Wang2017}, many-body physics~\cite{Arsenault.15,Carrasquilla2016a,Torlai.16,Chng.16,Aoki.16,Tubman.16,Carleo2017,LiuJunwei.17,Bukov.17},  classical \cite{Schoenholz.16} and quantum phase transitions \cite{WangLei.16,Nieuwenburg.17,ZhangYi.17}, and classification of quantum states~\cite{Broecker.16, Ma2017,Lu2017a}. While its full power on quantum control is yet to be revealed, in this work we are going to demonstrate that the dynamically corrected gates can be reliably constructed by training neural networks.

The remainder of the paper is organized as follows. In Sec.~\ref{sec:model} we present the model used in this work. After that, we show our results in Sec.~\ref{sec:res}. Results for pulses without noise correction are presented in Sec.~\ref{sec:resuncorr},  and those for noise-compensating pulses are shown in Sec.~\ref{sec:rescorr}. We conclude in Sec.~\ref{sec:conclusion}. In Appendix~\ref{appx:nnetwork}, we give a brief introduction on the neural network and supervised learning method.

\section{Model}\label{sec:model}

A singlet-triplet qubit, hosted by double quantum dots, is encoded in the $S_z=0$ subspace of two-spin states. The computational bases are $|0\rangle=|\mathrm{T_0}\rangle=({|\!\uparrow\downarrow\rangle}+{|\!\downarrow\uparrow\rangle})/\sqrt{2}$ and $|1\rangle=|\mathrm{S}\rangle=\left({|\!\uparrow\downarrow\rangle}-{|\!\downarrow\uparrow\rangle}\right)/\sqrt{2}$ \cite{Petta.05, Wang.12, Kestner.13,Wang.14a}. Under these bases, the control Hamiltonian can be written as
\begin{equation}
H(t)=\frac{J(t)}{2}\sigma_z+\frac{h}{2}\sigma_x.
\label{eq:Ham}
\end{equation}
Here, $h$ is the magnetic field gradient across the double quantum dots, constituting the $x$ rotation around the Bloch sphere. The exchange interaction $J(t)$ is associated with the $z$ rotation. Nevertheless, there are technical difficulties in varying $h$ during runtime, which means that the sole controllable parameter is the $z$-rotation rate---determined by the amplitude of $J$---which is then constrained by the energy level structure of the system to be non-negative. These special constraints have inspired construction of composite pulses in order to accomplish universal gates, with or without correction to environmental noises \cite{Wang.12, Kestner.13,Wang.14a}.
While constructing the composite pulses, the noise is assumed to be static, i.e. the nuclear noise adds an unknown term $\delta h$ so that $h\rightarrow h+\delta h$, and the charge noise causes $J(t)\rightarrow J(t)+\delta J(t)$ while $\delta J$ is represented as $\delta J(t)=J'[\varepsilon(t)]\delta \varepsilon$ where $\varepsilon$ is the detuning \cite{Petta.05}. The composite pulse sequences cancel both channels of noise to the leading order, and is only dependent on the desired rotation but not on the amplitude of noise.  The composite pulses are then numerically testified using randomized benchmarking simulations and are found to work reasonably well with noise that concentrates on low frequencies \cite{Wang.14a, Yang.16,Zhang.17}.

\begin{figure}
(a)\centering\includegraphics[width=0.5\columnwidth]{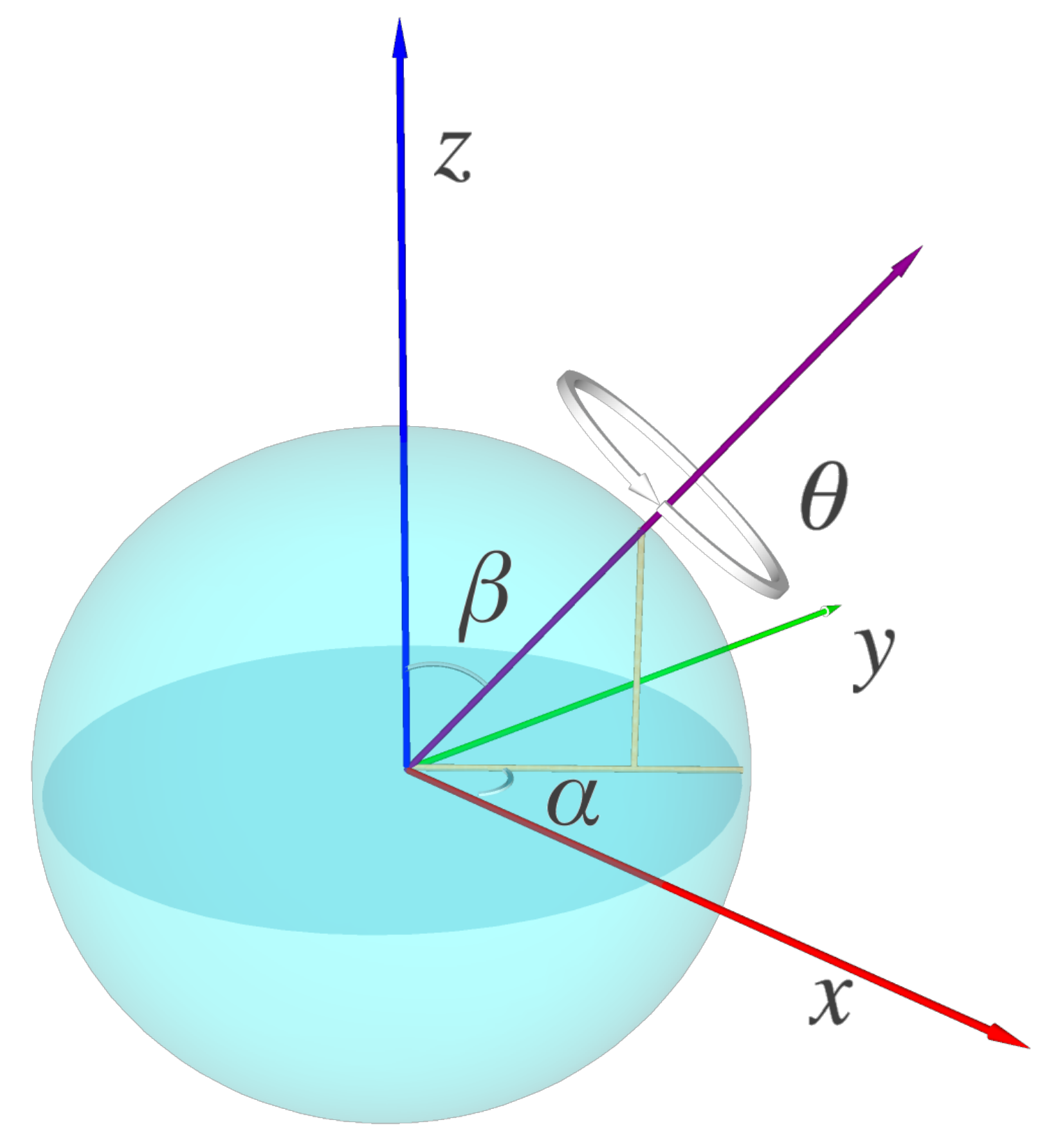}
\hspace{\fill}
(b)\centering\includegraphics[width=0.7\columnwidth]{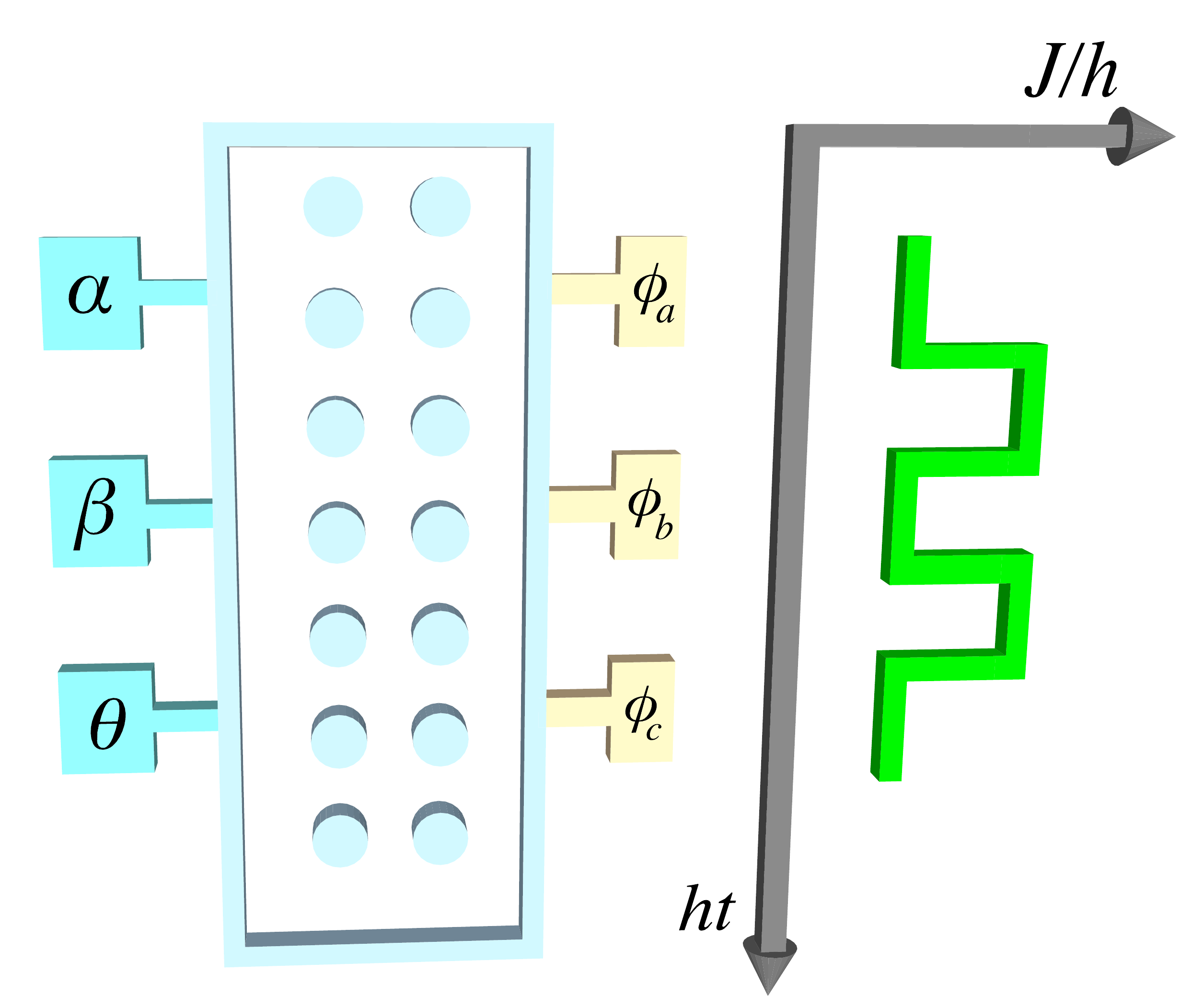}
\caption{(a) The Bloch sphere showing $\alpha$, $\beta$ and $\theta$ and the Cartesian coordinates. (b) Schematics of the neural network, along with the input $\{\alpha, \beta, \theta\}$, the output $\{\phi_a, \phi_b, \phi_c\}$ and the constructed composite pulse sequence. The neural network contains two layers with $N_n$ neurons each. $t$ is an arbitrary time unit.}
\label{fig:bloch}
\end{figure}

In this work, we explore the possibility of constructing composite pulses by supervised machine learning \cite{Mohri.12}. We first construct pulse sequences that are not immune to noises (usually called ``uncorrected'' or ``na\"ive'' ones \cite{Wang.12, Kestner.13,Wang.14a}), and then proceed to robust (``corrected'') ones.  Supervised machine learning requires a large amount of data as input, with which the various parameters (weights, biases) of the neural network is optimized, making the network capable of predicting the outcome from inputs that are not part of the training sets \cite{NielsenML}. More details on supervised learning are provided in Appendix~\ref{appx:nnetwork}.

\section{Results}\label{sec:res}

\subsection{Pulses without noise correction}\label{sec:resuncorr}

Even without noise correction, composite pulses are necessary for performing arbitrary rotations on the Bloch sphere, as the elementary rotations of a singlet-triplet qubit cover only the $x$ and $z$ axes. Arbitrary rotations must be decomposed into a sequence of such rotations, e.g. the $x$-$z$-$x$ decomposition \cite{NielsenChuang.00,Wang.14a}:
\begin{equation}
R(\hat{r},\theta)=e^{i\chi}R(\hat{x},\phi_a)R(\hat{z},\phi_b)R(\hat{x},\phi_c),\label{eq:xzxseq}
\end{equation}
where $\hat{r}$ is defined by a polar angle $\alpha$ and an azimuth angle $\beta$ ($\hat{r}=\cos\alpha\sin\beta\ \hat{i}+\sin\alpha\sin\beta\ \hat{j}+\cos\beta\ \hat{k}$) [cf. Fig.~\ref{fig:bloch}(a)], $\phi_{a,b,c}$ are auxiliary angles depending on the desired rotation, and $\chi$ is an unimportant overall phase. In practice the magnetic field gradient cannot be turned on and off during a given gate operation; the $z$ rotations must be further broken down \cite{NielsenChuang.00,Ramon.11,Wang.14a,Zhang.17}. For example, with the Hadamard-$x$-Hadamard sequence \cite{NielsenChuang.00} $R(\hat{z},\phi)=-R(\hat{x}+\hat{z},\pi)R(\hat{x},\phi)R(\hat{x}+\hat{z},\pi)$, Eq.~\eqref{eq:xzxseq} becomes a five-piece composite pulse \cite{Wang.14a},
\begin{equation}
\begin{split}
R(\hat{r},\theta)&=\\
e^{i\chi}R(\hat{x},&\phi_a)R(\hat{x}+\hat{z},\pi)R(\hat{x},\phi_b)R(\hat{x}+\hat{z},\pi)R(\hat{x},\phi_c).
\end{split}
\label{eq:5piece}
\end{equation}
The problem at hand is therefore to generate composite pulse sequence uniquely determined by $\{\phi_a, \phi_b, \phi_c\}$ from the desired rotation $R(\hat{r},\theta)$. 

Here, the input $\{\alpha, \beta, \theta\}$ and the output $\{\phi_a, \phi_b, \phi_c\}$ are mathematically related by a set of non-linear equations as follows:
\begin{widetext}
\begin{subequations}
\begin{align}
&\cos\frac{\theta}{2}-i\cos\beta\sin\frac{\theta}{2}=-\cos\frac{\phi_b}{2}\cos\frac{\phi_a+\phi_c}{2}+i\sin\frac{\phi_b}{2}\cos\frac{\phi_a-\phi_c}{2},\\
&\left(-i\cos\alpha-\sin\alpha\right)\sin\beta\sin\frac{\theta}{2}=-\sin\frac{\phi_b}{2}\sin\frac{\phi_a-\phi_c}{2}+i\cos\frac{\phi_b}{2}\sin\frac{\phi_a+\phi_c}{2},\\
&\left(-i\cos\alpha+\sin\alpha\right)\sin\beta\sin\frac{\theta}{2}=\sin\frac{\phi_b}{2}\sin\frac{\phi_a-\phi_c}{2}+i\cos\frac{\phi_b}{2}\sin\frac{\phi_a+\phi_c}{2}.
\end{align}
\label{eq:nonlinear1}
\end{subequations}
\end{widetext}
We are going to use a neural network to solve the problem. A schematics of the neural network together with the input, output and the five-piece composite pulse are shown in Fig.~\ref{fig:bloch}(b) (more details on the neural network are presented in Appendix~\ref{appx:nnetwork}). Note that we use $t$ to denote an arbitrary time unit throughout this paper.

\begin{figure}[t]
(a)\centering\includegraphics[width=0.75\columnwidth]{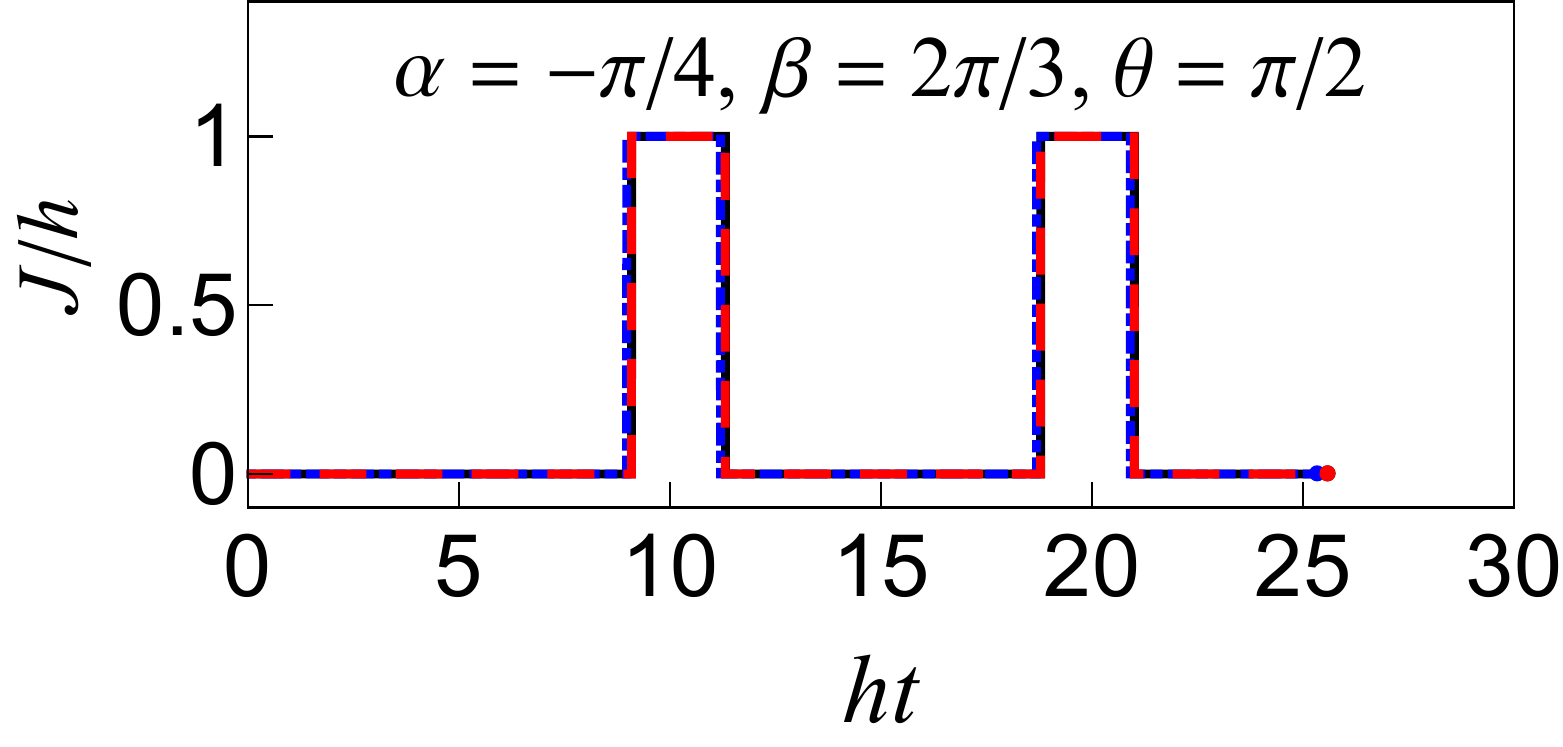}
\hspace{\fill}
(b)\centering\includegraphics[width=0.75\columnwidth]{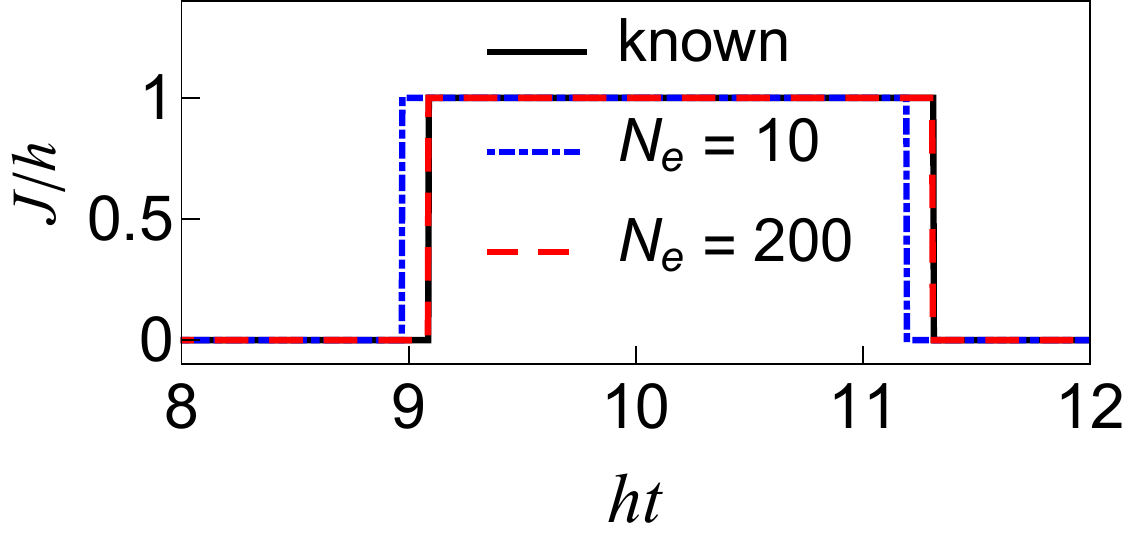}
\caption{Comparison of composite pulses predicted by the neural network and the known result for $\{\alpha, \beta, \theta\}$=$\{-\pi/4, 2\pi/3, \pi/2\}$. Panel (a) shows the entire pulse sequence and panel (b) is a zoom-in for the range $8\le ht \le12$. The black solid lines show the known results,  while the predictions from the network after $N_e=10$ and $N_e=200$ epochs are shown as blue/gray dash-dotted lines and red/gray dashed lines, respectively. $N_n=100$.}
\label{fig:uncorrpulse}
\end{figure}

Our neural network contains two layers with $N_n$ neurons each; the value of $N_n$ is adjusted during the learning process to generate optimal results. The training set is obtained as follows: 80 points are chosen non-uniformly for $\alpha\in[-\pi,0]$, excluding the following points due to discontinuity: $\alpha=-\pi,-3\pi/4,-\pi/2,-\pi/4,0$. To compensate this exclusion, training data are chosen more densely in the neighborhood of these points. On the other hand, data points are taken uniformly in sampling $\beta$ and $\theta$, where we have chosen 20 points for $\beta\in[0,\pi]$ and 40 points for $\theta\in[0,2\pi]$. (In this work, such sampling of the data points is sufficient for training the neural network. Alternative schemes of sampling, e.g. sampling over $\cos\alpha$ may also provide reasonably good training.) The training data set therefore contains $80\times20\times40=64000$ entries in total, and for each of them $\{\phi_a, \phi_b, \phi_c\}$ are calculated. These 64000 data points form our training set, which are then used to train the neural network for $N_e$ epochs \cite{NielsenML}. 
In practice, we have found that a learning rate  \cite{NielsenML} of 0.005 works best for our problem, and we will stick to it for all results presented in this paper. A list of default parameters of the neural network used in this work is presented in Table~I (see also Appendix~\ref{appx:nnetwork} for the meaning of the parameters). Predictions from the network are then compared to known results to benchmark its performance.

\begin{table}[]
  \label{tablenet}
  \begin{center}
    \caption{Default parameters of the neural network.}
    \begin{tabular}{p{6cm}rp{5cm}}
    \hline\hline
      Number of layers & 2\\
      Number of neurons in each layer $N_n$ & 100\\
      Size of the training data set $N_\mathrm{tr}$ & 64000\\
      Size of a data bin $N_b$ & 1\\
      Number of training epochs $N_e$ & 500 \\
      Activation function $f(z)$ & tanh($z$)\\
      Learning rate $\eta$ & 0.005\\
      \hline\hline
    \end{tabular}
  \end{center}
\end{table}

Figure~\ref{fig:uncorrpulse} shows a representative result for $\alpha=-\pi/4$, $\beta=2\pi/3$, $\theta=\pi/2$, with panel (b) a zoomed-in version of panel (a). The black lines show the known results obtained by solving the non-linear equations, while the predictions from the neural network after $N_e=10$ and $N_e=200$ epochs are shown as blue/gray dash-dotted lines and red/gray dashed lines, respectively. We see that after 10 epochs, the prediction from the neural network is already very close to the known result, with the difference only visible in the zoomed-in panel (b). After 200 epochs of training, the prediction from the neural network becomes essentially identical to the known result, demonstrating that the network is well trained and can make judicious predictions.

\begin{figure}[t]
\centering\includegraphics[width=\columnwidth]{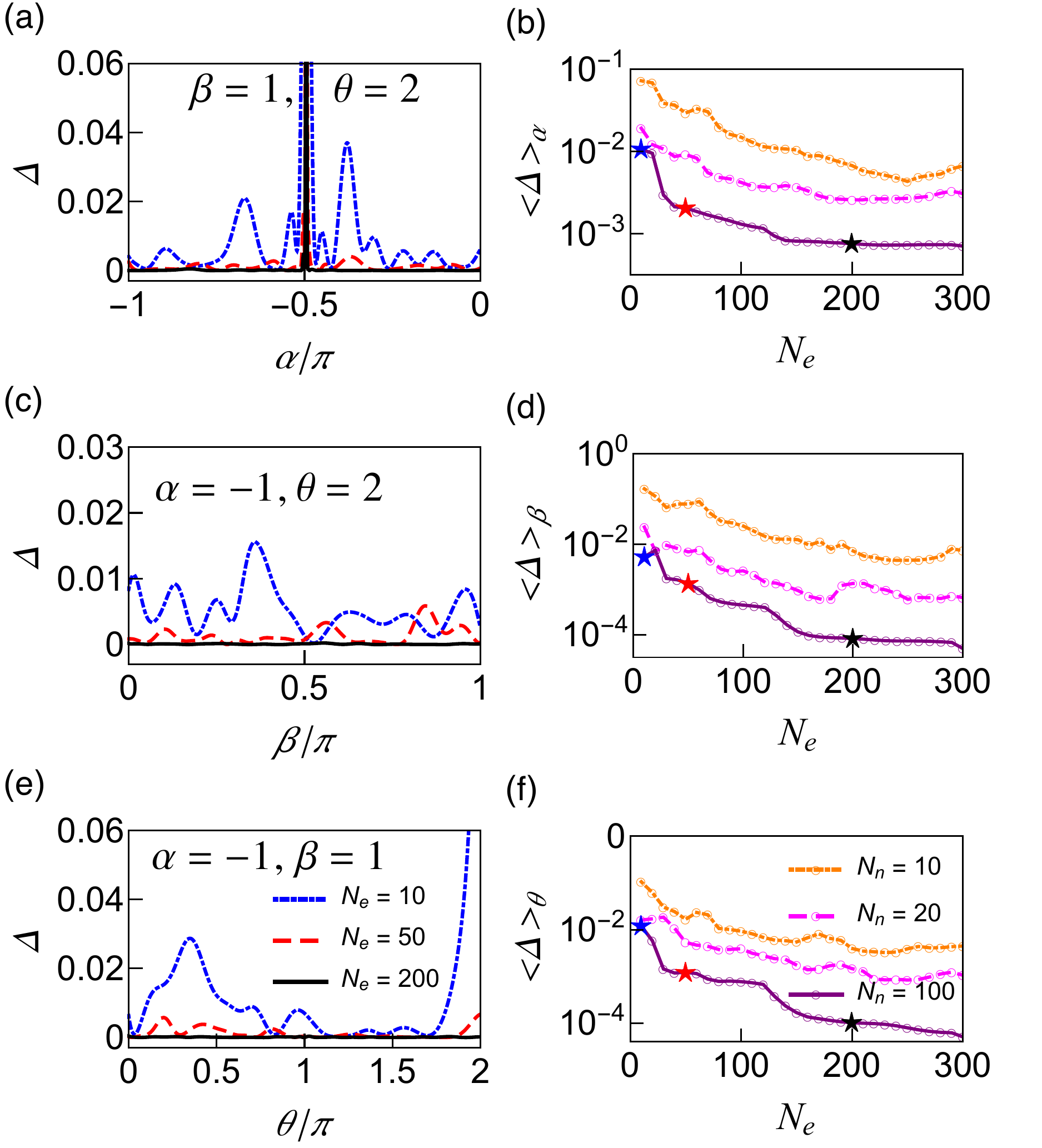}
\caption{The gate error $\Delta$ for sequences generated by the neural network. (a) $\Delta$ v.s.~$\alpha$ with other parameters fixed at $\beta=1$ rad and $\theta=2$ rad. (b) The gate error averaged for $\alpha\in[-\pi,0]$, $\langle\Delta\rangle_\alpha$ as a function of the number of epochs.
(c) $\Delta$ v.s.~$\beta$ with other parameters fixed at $\alpha=-1$ rad and $\theta=2$ rad. (d) The gate error averaged for $\beta\in[0,\pi]$, $\langle\Delta\rangle_\beta$ as a function of the number of epochs.
(e) $\Delta$ v.s.~$\theta$ with other parameters fixed at $\alpha=-1$ rad and $\beta=1$ rad. (f) The gate error averaged for $\theta\in[0,2\pi]$, $\langle\Delta\rangle_\theta$ as a function of the number of epochs. For (a), (c), (e), the blue/gray dash-dotted lines, red/gray dashed lines and the black solid lines show results generated by $N_e=10, 50$ and $200$ epochs respectively, from a neural network with $N_n=100$. The average gate errors for the three curves in each panel are shown in their corresponding right column panels as stars with the same color.
For (b), (d), (f), the orange dash-dotted lines, magenta dashed lines and purple solid lines represent results from neural networks with $N_n=10, 20$ and $100$, respectively.
We note that the sets of parameters used in this figure are not in the training set.}
\label{fig:epoch}
\end{figure}

To further investigate the performance of the neural network, we define the gate error as
\begin{equation}
\Delta \equiv 1-\overline{\left|\langle\psi_i|V^\dagger U|\psi_i\rangle\right|^2},
\end{equation}
where $V$ is the desired operation, $U$ is the operator for the actual evolution, and their overlap is averaged over initial states $\psi_i$ distributed uniformly around the Bloch sphere. We plot $\Delta$ as functions of various parameters in Fig.~\ref{fig:epoch}. Fig.~\ref{fig:epoch}(a), (c) and (e) show $\Delta$ as functions of $\alpha$, $\beta$ and $\theta$ respectively with remaining angles fixed as indicated. The blue/gray dash-dotted lines, red/gray dashed lines and the black solid lines show the results for $N_e=10$, 50 and 200 respectively. It is clear that after $N_e=10$ epochs, the error is still large, at the order of about a few percent. However, after $N_e=50$ epochs the error are suppressed below 1\%, and $N_e=200$ epochs are sufficient to reduce the error to about $10^{-4}$. Nevertheless, for certain special angles such as $\alpha=-\pi/2$ (the ``singular'' point) as indicated in Fig.~\ref{fig:epoch}(a), the neural network would fail to predict the correct sequence. For small $N_e$, the failure seems to cover a small neighborhood around the singular point, but as $N_e$ is increased to 200 such failure is well confined at the singular point alone. While the reason why this happens is not fully understood, we suspect this originates from the discontinuity of relevant solutions to the non-linear equations.

The right column of Fig.~\ref{fig:epoch} shows the gate error averaged over their respective angles, as functions of the number of epochs $N_e$ for three different values of $N_n$ as indicated.  We see that increasing the number of neurons has a significant impact on the learning process: for $N_n=10$ the gate error saturates at about 1\% after sufficient training, and this value drops substantially as $N_n$ is increased. For $N_n=100$, the best gate error is lower than $10^{-4}$ for $\langle\Delta\rangle_\beta$ and $\langle\Delta\rangle_\theta$, and is about $10^{-3}$ for
$\langle\Delta\rangle_\alpha$ due to the singularity at $\alpha=-\pi/2$. These results suggest that $N_n=100$ neurons per layer is sufficient to produce composite pulses with errors less than the fault-tolerant threshold, with the exception of singular points.

\begin{figure}[t]
(a)\centering\includegraphics[width=0.7\columnwidth]{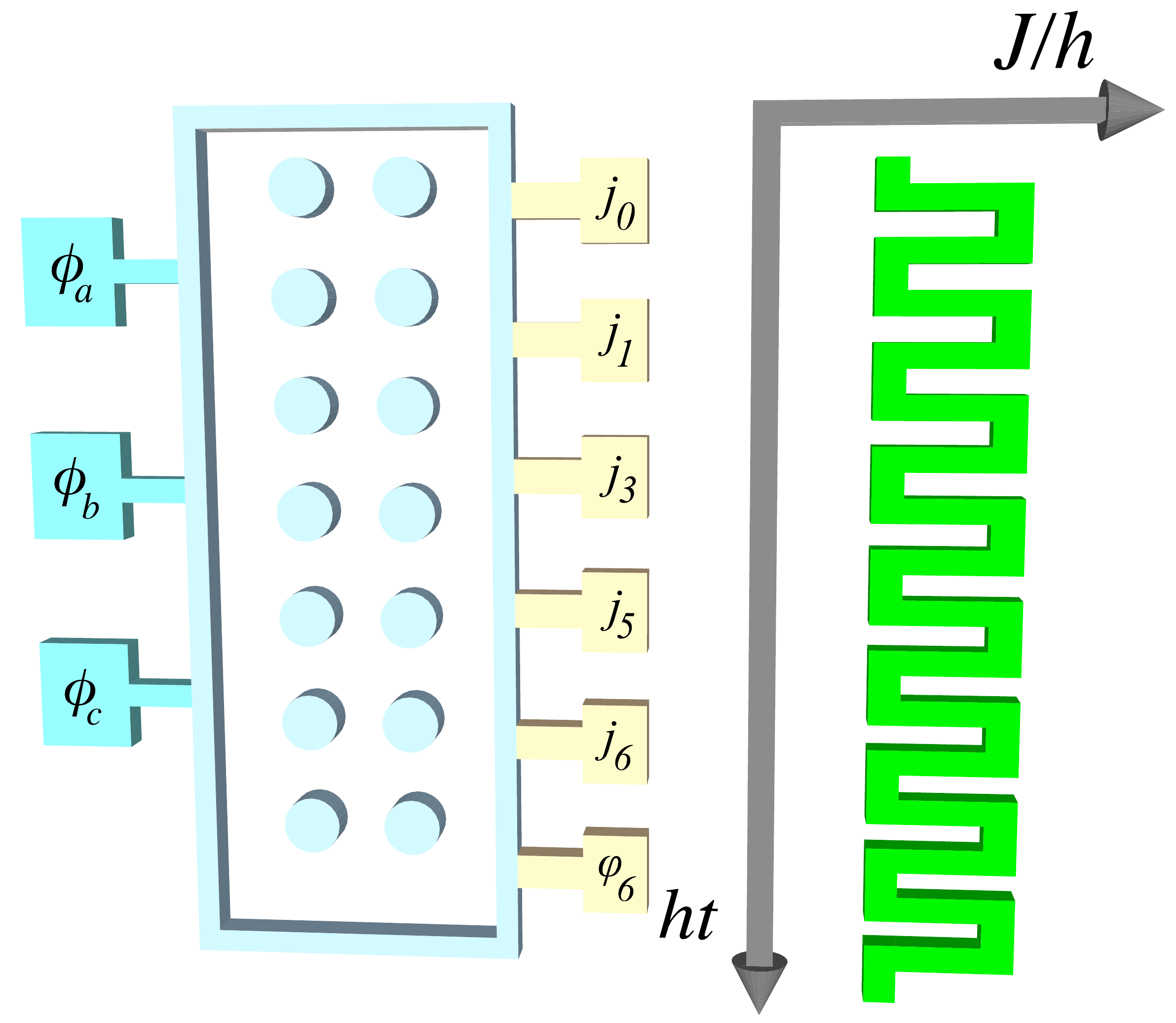}
\ \\
\ \\
\ \\
(b)\centering\includegraphics[width=0.9\columnwidth]{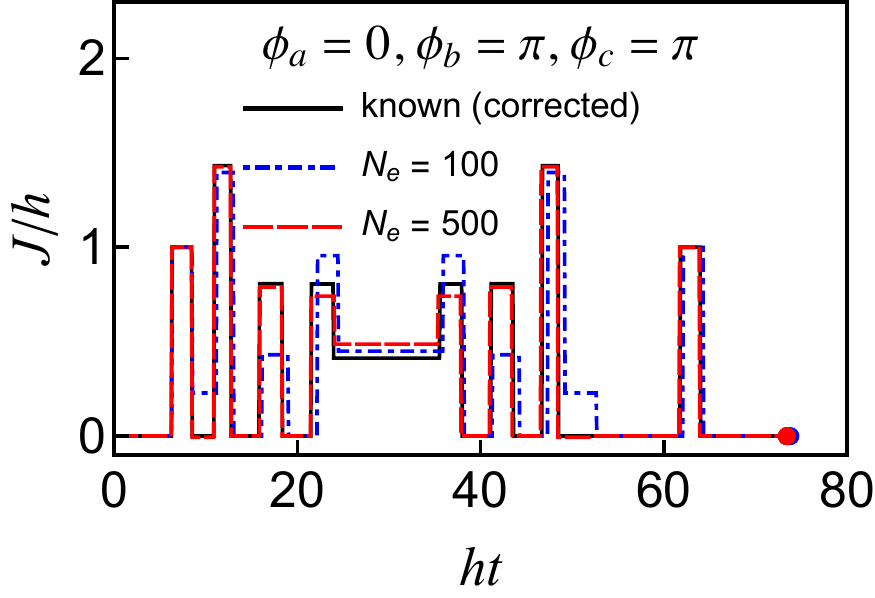}
\caption{(a) 
Schematics of the neural network used, along with the input $\{\phi_a, \phi_b, \phi_c\}$, the output $\{j_i,\varphi_6\}$ and the constructed composite pulse sequence, which is uniquely determined by $\{j_i,\varphi_6\}$ $(i=0,1,3,5,6)$. The neural network contains two layers with $N_n$ neurons each. (b) Comparison of robust composite pulse sequences for a rotation defined by $\{\phi_a, \phi_b, \phi_c\}=\{0, \pi, \pi\}$. Black solid line: the known sequence correcting noise. Blue/gray dash-dotted line: the result generated by the neural network after $N_e=100$ epochs of training. Red/gray dashed line: the result generated after $N_e=500$ epochs of training.}
\label{fig:corrpulseschem}
\end{figure}

\subsection{Noise-correcting pulses}\label{sec:rescorr}

Next, we proceed to composite pulse sequences that can correct hyperfine and charge noise in a singlet-triplet qubit. The basic idea of {\sc supcode} is to invoke the self-compensation of noise: we first apply a ``na\"ive rotation", which is not immune to noise; then we supplement it by a lengthy identity which is carefully engineered such that the errors arising from the identity would exactly cancel those derived in the na\"ive rotation to the leading order, thereby achieving noise-robustness for the entire sequence. Such robustness of the ``corrected sequence'' is therefore accomplished at a cost of a prolonged gate time. For a general rotation already decomposed as Eq.~\ref{eq:xzxseq}, the corrected sequence is
\begin{align}
\begin{split}
&
U(0,\phi_a)
U(1,\pi)
U(j_6,\pi-\varphi_6)
U(j_5,\pi)
U(j_4=0,\pi)\\
&\times
U(j_3,\pi)
U(j_2=0,\pi)
U(j_1,\pi)
U(j_0,4\pi)
U(j_1,\pi)\\
&\times
U(j_2=0,\pi)
U(j_3,\pi)
U(j_4=0,\pi)
U(j_5,\pi)\\
&\times
U(j_6,\pi+\varphi_6)
U(0,\phi_b)
U(1,\pi)
U(0,\phi_c),
\end{split}\label{eq:FivePieceSeq}
\end{align}
which is uniquely determined by five exchange parameters $j_0$, $j_1$, $j_3$, $j_5$, $j_6$, and one angular parameter $\varphi_6$ (cf. Eq.~(32) in Ref.~\onlinecite{Wang.14a}). The problem then boils down to one with input $\{\phi_a, \phi_b, \phi_c\}$ and output $\{j_0, j_1, j_3, j_5, j_6, \varphi_6\}$. 
Here, the evolution operator for a ``single-piece'' control is
\begin{align}
&U\left(J,\phi\right) \equiv \exp{\left[-i\left(\frac{h+\delta h}{2} \sigma_x + \frac{J+\delta J}{2} \sigma_z\right)\frac{\phi}{\sqrt{h^2+J^2}}\right]},\label{eq:UJphi}
\end{align}
with $h=1$ as the energy unit. Fig.~\ref{fig:corrpulseschem}(a) shows a schematics of this setup.

\begin{figure}[t]
\centering\includegraphics[width=1\columnwidth]{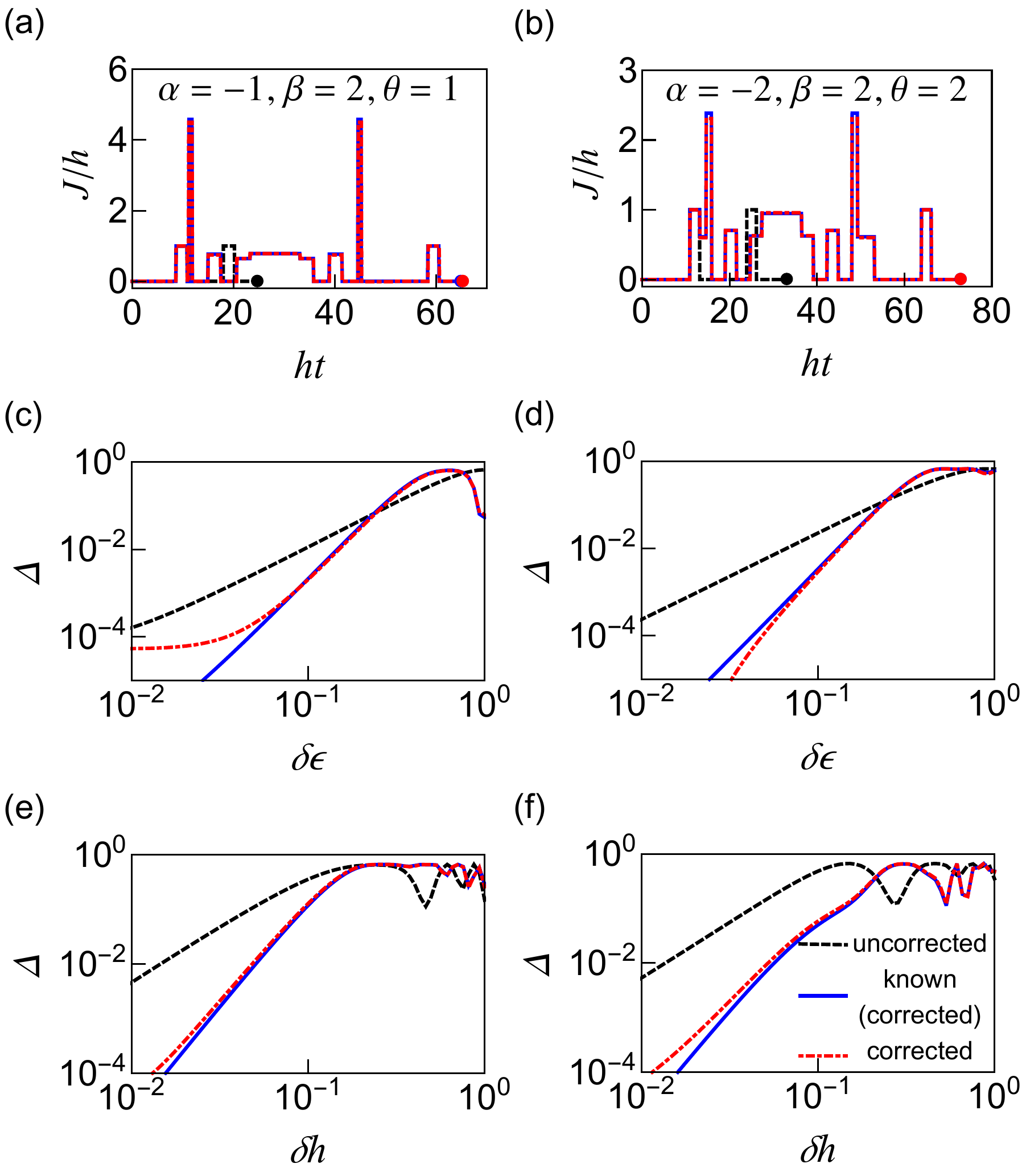}
\caption{Comparison of the pulse profiles and gate error of the na\"ive and corrected sequences. Left column (a), (c), (e): results for a rotation defined by $\alpha=-1$ rad, $\beta=2$ rad, and $\theta=1$ rad; right column (b), (d), (f): results for a rotation defined by $\alpha=-2$ rad, $\beta=2$ rad, and $\theta=2$ rad.
(a), (b): Comparison of na\"ive and corrected pulse sequences. Na\"ive pulse sequences as predicted by the neural network are shown as the black dashed line (``uncorrected'') and the corrected ones generated by the network are shown as red/gray dash-dotted lines (``corrected''). The known corrected pulse sequences are shown as the blue/gray solid lines (``known (corrected)''). (c), (d): Comparison of the gate errors as functions of the charge noise ($\delta\epsilon$) with the nuclear noise set to zero. (e), (f): Comparison of the gate errors as functions of the nuclear noise ($\delta h$) with the charge noise set to zero. Parameters: $N_n=100$. For results shown in the left column, $N_e=200$; and $N_e=500$ for those shown in the right column.}
\label{fig:corrpulse}
\end{figure}

We use the same $64000$ rotations as in the previous part as the training data, with $\{\phi_a, \phi_b, \phi_c\}$ (with $2\pi$ subtracted as appropriate)  as inputs and calculated $\{j_0, j_1, j_3, j_5, j_6, \varphi_6\}$ as outputs. Technically it is easier to train the neural network if all the angles are normalized by $2\pi$ and the exchange parameters by $J_{\rm max}$, a maximum value found in the sequence. Fig.~\ref{fig:corrpulseschem}(b) compares the robust composite pulse sequence generated by the neural network to the known one,  for a rotation defined by $\{\phi_a, \phi_b, \phi_c\}=\{0, \pi, \pi\}$. Black solid line shows the known sequence. After $N_e=100$ epochs of training, the neural network is able to predict the basic shape of the pulse sequence as shown by the blue/gray dash-dotted line. After $N_e=500$ epochs,  the pulse sequence generated by the neural network is already very close to the known one, with only small differences, as seen from the red/gray dashed line. Despite these small differences,   the performance in correcting noise of the pulse sequences generated by the neural network is about the same as compared to those found in the literature, as shall be demonstrated immediately below.

In order to reveal the robustness of the pulse sequences generated by the neural network, we compare the corresponding gate error as functions of the charge noise and the nuclear noise in Fig.~\ref{fig:corrpulse}. The left column [panels (a), (c), (e)] shows results for a rotation defined by $\alpha=-1$ rad, $\beta=2$ rad, and $\theta=1$ rad, while the right column [panels (b), (d), (f)] shows those defined by $\alpha=-2$ rad, $\beta=2$ rad, and $\theta=2$ rad. Three kinds of pulse sequences are compared: the na\"ive sequence (``uncorrected'', black dashed lines) generated by the neural network, the corrected sequence (``corrected'', red/gray dash-dotted lines) generated by the network, and the theoretically known corrected sequence (``known (corrected)'', blue/gray solid lines). Their pulse profiles are compared in panels (a) and (b). In Fig.~\ref{fig:corrpulse}(c) and (d), we set the nuclear noise as zero and plot the dependence of the gate error of the three kind of sequences on the charge noise $\delta\epsilon$. We see that while the results for the corrected sequences generated by the neural network do not exactly match those for the known sequences, the two sets are very similar. In Fig.~\ref{fig:corrpulse}(c), large deviation only occurs for $\Delta<10^{-4}$, which is not important for the purpose of quantum error-correction. Similar conclusions may be drawn from Fig.~\ref{fig:corrpulse}(e) and (f), where the gate error as functions of the nuclear noise $\delta h$ with the charge noise fixed to zero are shown. Therefore, the neural network has demonstrated its excellent power in generating composite pulse sequences for operation of a spin qubit, either uncorrected or corrected against noise.

We note that the training of the neural network typically takes a long time (and in this case, about 20 to 30 hours on a 3.1GHz CPU with 8 GB memory). However, once it is trained, it is much more efficient than the original method of finding real and non-negative solutions to a set of nonlinear equations  because an output is generated straight from the network while the solution-finding usually requires multiple iterations. For a typical data point we have, it takes $\sim 0.01$ sec for the neural network to generate a solution, and $\sim 4$ sec for the traditional method of solving nonlinear equations.

\section{Conclusion}\label{sec:conclusion}

In conclusion, we have demonstrated that the composite pulse sequences theoretically proposed for operating a spin qubit, with or without noise robustness, can be generated by a trained neural network. While simple composite pulse sequences can be implemented experimentally, the more complicated error-correcting sequences as used here are still too complicated to be realized, so it would be difficult to observe the difference in performances between the theoretically proposed sequences and those generated by neural networks. Moreover, systematic errors in performing quantum control can be significant, but these have not been considered in the present theoretical framework. We hope further development at the interface of machine learning and the robust quantum control can simplify the pulse sequences so they are easier to implement, and can include more realistic error sources so they perform better when realized in experiments. We believe we are  only at the beginning of revealing the full power of machine learning in controlling quantum systems, and hope that our results would inspire further research in applying the machine learning technique to quantum control of spin qubits.

This work is supported by the 
Research Grants Council of the Hong Kong Special Administrative Region, China (Nos.~CityU 21300116, CityU 11303617), the National Natural Science Foundation of China (Nos.~11604277, 11405093), the Guangdong Innovative and Entrepreneurial Research Team Program (No.~2016ZT06D348), and the Science, Technology and Innovation Commission of Shenzhen Municipality (ZDSYS201703031659262, JCYJ20170412152620376).

%\onecolumngrid

\appendix
\setcounter{equation}{0}

\section{A brief introduction of the neural network and supervised learning}\label{appx:nnetwork}

In this section we give a very brief introduction of the neural network and supervised learning. This section largely follows \cite{NielsenML}.

\begin{figure}[t]
\centering\includegraphics[width=1\columnwidth]{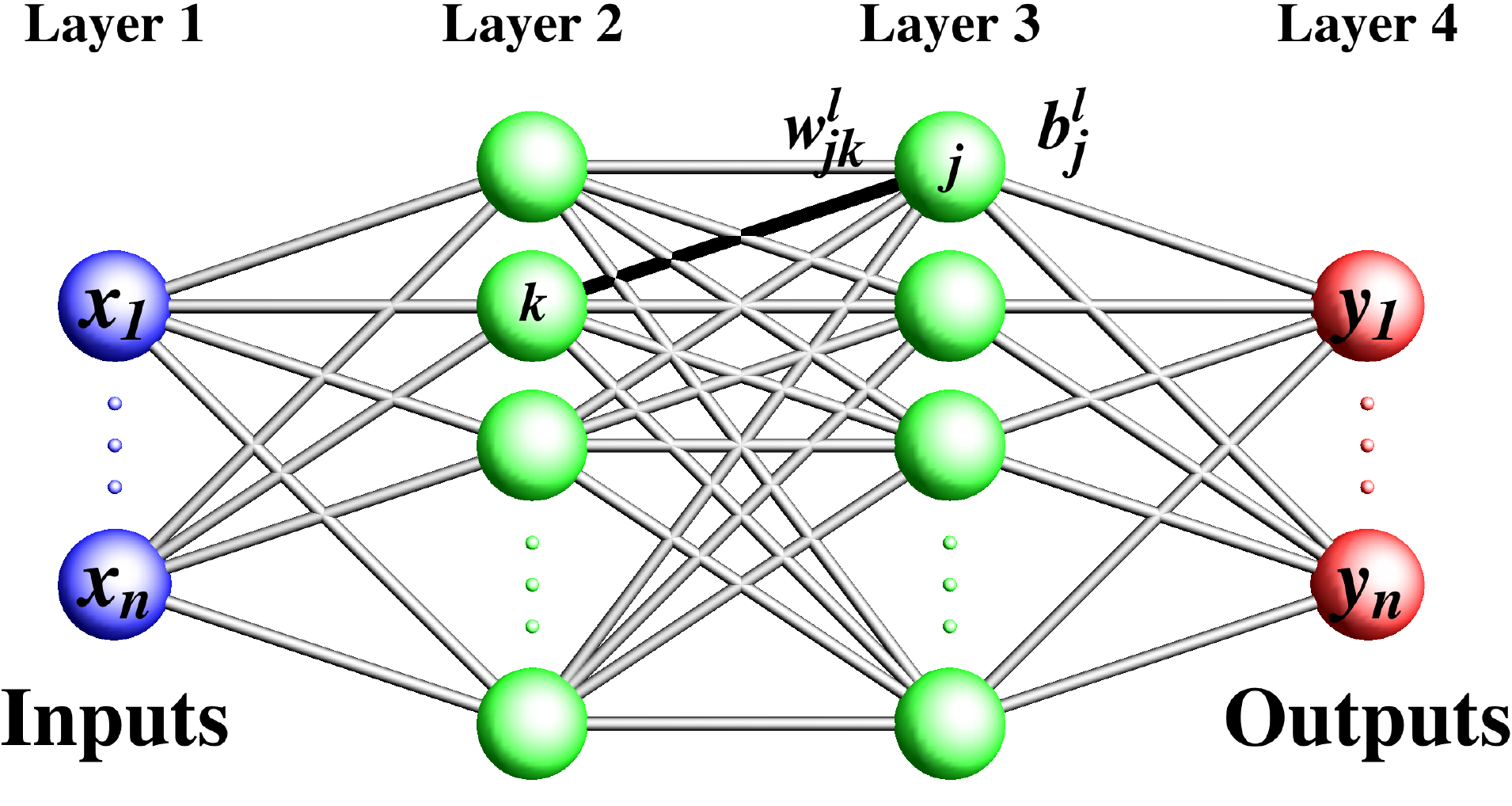}
\caption{Schematics of a double-layer neural network. We call the inputs ``layer 1'', the two layers of neurons ``layer 2'' (left) and ``layer 3'' (right), and the outputs ``layer 4''. }
\label{fig:nnetwork}
\end{figure}

Fig.~\ref{fig:nnetwork} shows a schematic diagram of a double-layer neural network. Each neuron carries a ``bias'' and a set of ``weights'', which are used to calculate its outputs to the next layer with the inputs from the previous layer. The output of the $j^{\rm th}$ neuron in the $l^{\rm th}$ layer (also called the ``activation''), $a_j^l$, is related to the inputs from the $k^{\rm th}$ neuron in the previous layer, $a_k^{l-1}$, by
\begin{equation}
a_j^l=f\left(\sum_k w_{jk}^la_k^{l-1}+b_j^l\right),
\end{equation}
where $w_{jk}^l$ is the weight of the $j^{\rm th}$ neuron in the $l^{\rm th}$ layer corresponding to the input from the $k^{\rm th}$ neuron in the $(l-1)^{\rm th}$ layer (typically confined between 0 and 1), $b_j^l$ is the bias of the $j^{\rm th}$ neuron in the $l^{\rm th}$ layer. For the convenience of later discussions, we define a ``weighted input'' $z_j^l=\sum_k w_{jk}^la_k^{l-1}+b_j^l$ so that $a_j^l=f(z_j^l)$. Here, $f(z)$ is  the ``activation function'' dependent on the specific type of neurons chosen.

In the simplest case, we could take the neurons as ``perceptrons'', i.e. $f(z)=\Theta(z)$, where $\Theta(z)$ is the Heaviside step function giving 1 for $x>0$ and 0 otherwise.  In this case, the neuron is more likely to ``fire'' (i.e. $f(z)=1$) when it has a positive bias or when the inputs from the previous layer are mostly positive. In this work, we use the ``tan-sigmoid'' neurons, i.e $f(z)=\tanh(z)$. This kind of ``smoothed-out'' step function turns out to be the key for the network to efficiently learn.

\begin{figure}[t]
\centering\includegraphics[width=1\columnwidth]{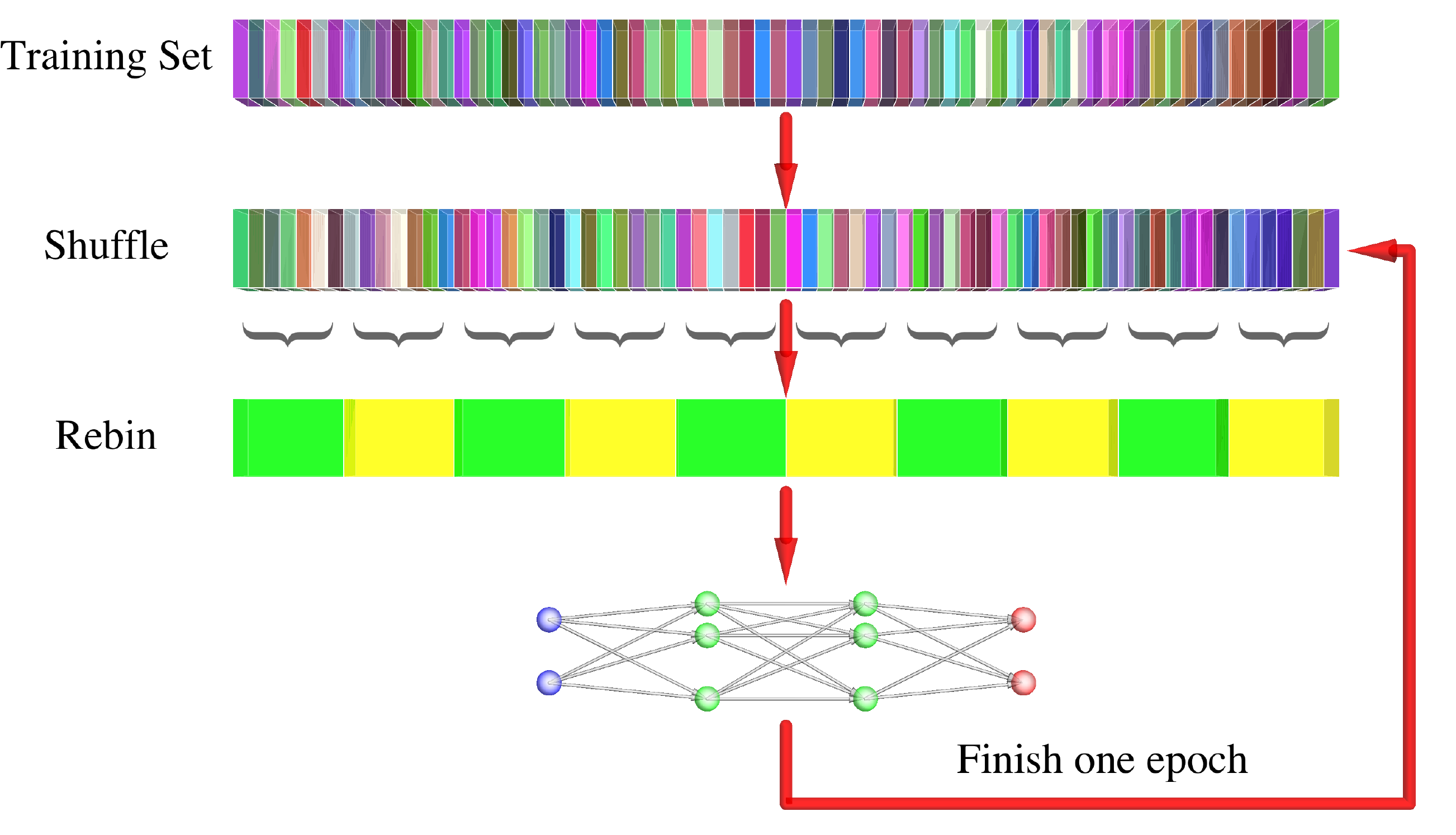}
\caption{Schematics of supervised learning. We start with a training set with different data entries represented by different colors. In an epoch of training, the training set is first shuffled, and every $N_b$ entries are combined (averaged) into data bins. These data are fed to the neural network so that it can adjust weights and bias with the known inputs and outputs. Then the training set is shuffled, re-binned again and another epoch of training is exercised. One repeats this process until the predicted outputs agree with known ones to satisfactory levels.}
\label{fig:epoch2}
\end{figure}

During the course of supervised learning, a neural network ``learns'' from a set of data with known inputs and outputs (``supervised''). A properly trained network is vested with power to predict unknown outputs from some given inputs.  The learning process is essentially one that keeps modifying the weights and bias of all neurons such that the prediction from the network fits the training set best.

Fig.~\ref{fig:epoch2} shows schematically one epoch of training for a general problem.   We start with a training set with $N_\mathrm{tr}$ different data entries represented by different colors. The training set is first shuffled so that data are randomly reordered. Then, every $N_b$ entries are combined into data bins, i.e. the average inputs and outputs of all data in a bin are used to train the network. $N_b$ is therefore called the bin size. For our specific problem of generating composite pulse sequences, we have chosen $N_b=1$ as there is essentially no noise in the training data. Nevertheless, re-binning is commonly performed for other problems involving noises in the training data because too many data points with fluctuations lead to ``overfitting'' so that the network confuses itself with noisy details of the data. Averaging data in a bin will smooth out the fluctuations, facilitating the learning process.

These $N_\mathrm{tr}/N_b$ data points are then fed to the neural network. The neural network first makes a prediction of outputs from known inputs in these data. Then, these predictions are compared to the known outputs and the differences are ``back-propagated'' throughout the network to modify the weights and bias of various neurons with the hope that it fits the data better in the next epoch. To understand the back-propagation algorithm, we define a cost function for one training example $x$ (which has a quadratic form) as
\begin{equation}
C_x=\frac{1}{2}\sum_j\left(y_j-a_j^L\right)^2,
\end{equation}
where $y_j$ is the desired 
(known) output on the $j^{\rm th}$ neuron in the output (last) layer, $L$ denotes the total number of layers in the network so $a_j^L$ is the actual output at the same place. The cost function for the entire training data set is an average over all training examples:
 \begin{equation}
C=\frac{1}{N_\mathrm{tr}}\sum_xC_x.
\end{equation}
The goal of the back-propagation is to find partial derivatives (gradients) $\partial C/\partial w_{jk}^l$ and $\partial C/\partial b_j$ so that the neurons know how their weights and biases are to be modified in order to fit the data better, reducing the cost function. Using the weighted input defined above, we define a more accessible ``error'' $\delta_j^l$ of the $j^{\rm th}$ neuron in the $l^{\rm th}$ layer as
 \begin{equation}
\delta_j^l\equiv\frac{\partial C}{\partial z_j^l},
\end{equation}
and one may readily find that  $\partial C/\partial b_j=\delta_j^l$ and
$\partial C/\partial w_{jk}^l=a_k^{l-1}\delta_j^l$. 

The remaining task is to find $\delta_j^l$. We start with the output layer, where
\begin{equation}
\delta_j^L=\frac{\partial C}{\partial z_j^L}=\sum_k\frac{\partial C}{\partial a_k^L}\frac{\partial a_k^L}{\partial z_j^L}.
\label{eq:delta1}
\end{equation}
Here, $\partial a_k^L/\partial z_j^L$ must be nonzero only if $k=j$. So Eq.~\eqref{eq:delta1} can be rewritten as
\begin{equation}
\delta_j^L=\frac{\partial C}{\partial a_j^L}\frac{\partial a_j^L}{\partial z_j^L}=\frac{\partial C}{\partial a_j^L}f'(z_j^L).
\label{eq:delta2}
\end{equation}
Note that either of the two terms in the r.h.s. of Eq.~\eqref{eq:delta2} can be easily obtained.

We now ``back-propagate'' the error to previous layers, i.e. we are going to relate $\delta_j^{l}$ to $\delta_j^{l+1}$. Note that
\begin{equation}
\delta_j^l=\frac{\partial C}{\partial z_j^l}=\sum_k\frac{\partial C}{\partial z_k^{l+1}}\frac{\partial z_k^{l+1}}{\partial z_j^l}=\sum_k\delta_k^{l+1}\frac{\partial z_k^{l+1}}{\partial z_j^l}.
\label{eq:delta3}
\end{equation}
We also have 
\begin{equation}
z_k^{l+1}=\sum_j w_{kj}^{l+1}f(z_j^l)+b_k^{l+1}
\end{equation}
so 
\begin{equation}
\frac{\partial z_k^{l+1}}{\partial z_j^l}=w_{kj}^{l+1}f'(z_j^l).
\label{eq:delta4}
\end{equation}
Combining Eqs.~\eqref{eq:delta3} and \eqref{eq:delta4}, we have
\begin{equation}
\delta_j^l=f'(z_j^l)\sum_kw_{kj}^{l+1}\delta_k^{l+1}.
\label{eq:delta5}
\end{equation}
Eqs.~\eqref{eq:delta2} and \eqref{eq:delta5} are sufficient to find all $\delta_j^l$, with which the partial derivatives of the cost function with respect to the weights and biases can be obtained and consequently the network can be updated.

We note that this is essentially a gradient descent method and one has to maintain relatively small steps of change by discounting the change to the parameters, namely 
\begin{subequations}
\begin{align}
	w_{jk}^l &\rightarrow \widetilde{w}_{jk}^{l}=w_{jk}^l-\eta\partial C/\partial w_{jk}^l,\\
	b_j &\rightarrow \widetilde{b}_j=b_j-\eta\partial C/\partial b_j.
\end{align}		
\end{subequations}
Here, the discount is called the ``learning rate'', denoted by $\eta$. When $\eta$ is too small the network learns slowly, but if $\eta$ is too large the updating to the network becomes unstable. In order to achieve satisfactory training results, an appropriate $\eta$ should be chosen. After the back-propagation process is done, an epoch is finished and one starts another epoch by reshuffling the training set.

\begin{figure*}[t]
\centering\includegraphics[width=1.8\columnwidth]{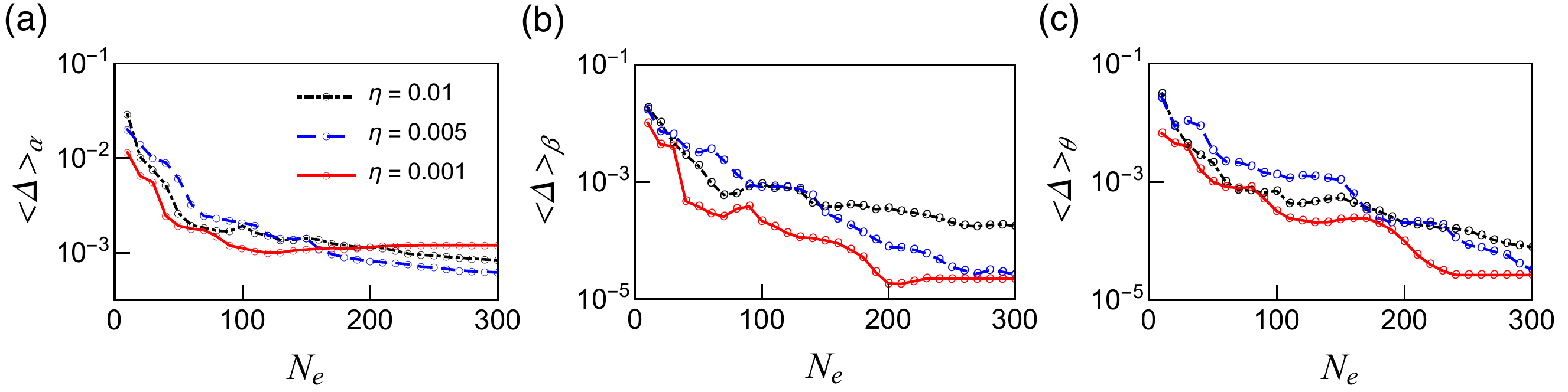}
\caption{The average gate error v.s. number of epochs in a training using gates that are not immune to noise. (a) The gate error averaged for $\alpha\in[-\pi,0]$, $\langle\Delta\rangle_\alpha$, as a function of the number of epochs. Other parameters: $\beta=1$ rad and $\theta=2$ rad.
(b) The gate error averaged for $\beta\in[0,\pi]$, $\langle\Delta\rangle_\beta$, as a function of the number of epochs. Other parameters: $\alpha=-1$ rad and $\theta=2$ rad.
 (f) The gate error averaged for $\theta\in[0,2\pi]$, $\langle\Delta\rangle_\theta$, as a function of the number of epochs. Other parameters: $\alpha=-1$ rad and $\beta=1$ rad.
The black dash-dotted lines, blue/gray dashed lines and red/gray solid lines are results for the learning rate $\eta=0.01$, 0.005, and 0.001 respectively. The results are calculated from a double-layer neural network with $N_n=50$.}
\label{fig:learnrate}
\end{figure*}

Figure \ref{fig:learnrate} shows the learning process in a training using gates that are not immune to noise. The three panels show the average gate errors for different angles, $\langle\Delta\rangle_\alpha$, $\langle\Delta\rangle_\beta$, and $\langle\Delta\rangle_\theta$ v.s. number of epochs $N_e$ respectively. In all cases, the errors are large at the beginning of the training (small $N_e$). As $N_e$ increases, the errors drop. One may then declare that the network is trained once the error is consistently below certain threshold. Fig.~\ref{fig:learnrate} also gives data for different learning rates. One may expect that for a large learning rate, the error drops quickly but there may be more fluctuations since the back-propagation of the output mismatch may be large enough to bring the network away from a local minimum in the parameter space, while for a small learning rate, a steady but possibly slow drop to a desired precision is expected. However, the results are not always the case since the training process involves many random factors. One should just choose a learning rate which gives the best compromise between the resource cost and the training quality desired. Other parameters of the network, e.g. the number of neurons in each layer $N_n$ and the bin size are determined with similar considerations.

%\bibliography{refs_robustML}

%merlin.mbs apsrev4-1.bst 2010-07-25 4.21a (PWD, AO, DPC) hacked
%Control: key (0)
%Control: author (72) initials jnrlst
%Control: editor formatted (1) identically to author
%Control: production of article title (-1) disabled
%Control: page (0) single
%Control: year (1) truncated
%Control: production of eprint (0) enabled
%

\end{document}